\documentstyle[aps,preprint,floats,tighten]{revtex}

\begin{document}
\draft
% \twocolumn[\hsize\textwidth\columnwidth\hsize\csname @twocolumnfalse\endcsname
\preprint{\vbox{\hbox{CU-TP-763}
                \hbox{CAL-612}
                \hbox{UM-AC 96-08}
                \hbox{hep-ph/9609370}
}}

\title{Indirect Detection of a Light Higgsino Motivated by
Collider Data}

\author{Katherine Freese\footnote{freese@mich.physics.lsa.umich.edu}}
\address{Randall Physics Laboratory, University of Michigan, Ann
Arbor, Michigan 48109-1120}
\author{Marc Kamionkowski\footnote{kamion@phys.columbia.edu}}
\address{Department of Physics, Columbia University, 538 West
120th St., New York, New York~~10027}

\date{September 1996}
\maketitle

\begin{abstract}
Kane and Wells recently argued that collider data point to a
Higgsino-like lightest supersymmetric partner which would
explain the dark matter in our Galactic halo.  They discuss
direct detection of such dark-matter particles in laboratory
detectors.  Here, we argue that such a particle, if it is indeed
the dark matter, might alternatively be accessible in
experiments which search for energetic neutrinos from
dark-matter annihilation in the Sun.   We provide accurate analytic
estimates for the rates which take into account all relevant physical
effects.  Currently, the predicted signal falls roughly one to three
orders of magnitude  below experimental bounds, depending on the mass
and coupling of the particle; however, detectors such as MACRO,
super-Kamiokande, and AMANDA will continue to take data and should be
able to rule out or confirm an interesting portion
of the possible mass range for such a
dark-matter particle within the next five years. 

\end{abstract}

% ]

\def\newpage{\vfill\eject}
\def\pp{\parshape 2 0.0truecm 16.25truecm 2truecm 14.25truecm}
\def\msun{\mass_\odot}
\def\sun{{\odot}}
\def\dwig{ {\widetilde d} } 
\def\lwig{ {\widetilde \ell} } 
\def\nro{ n_{R0}  } 
\def\fun#1#2{\lower3.6pt\vbox{\baselineskip0pt\lineskip.9pt
  \ialign{$\mathsurround=0pt#1\hfil##\hfil$\crcr#2\crcr\sim\crcr}}}
\def\la{\mathrel{\mathpalette\fun <}}
\def\ga{\mathrel{\mathpalette\fun >}}
\def\slashchar#1{\setbox0=\hbox{$#1$}           % set a box for #1
   \dimen0=\wd0                                 % and get its size 
   \setbox1=\hbox{/} \dimen1=\wd1               % get size of /
   \ifdim\dimen0>\dimen1                        % #1 is bigger 
      \rlap{\hbox to \dimen0{\hfil/\hfil}}      % so center / in box 
      #1                                        % and print #1
   \else					% /  is bigger 
      \rlap{\hbox to \dimen1{\hfil$#1$\hfil}}   % so center #1
      /                                         % and print /
   \fi}                                         %
\def\VEV#1{\left\langle #1\right\rangle}
\def\Nzsq{\VEV{Nz^2}}

\section{INTRODUCTION}

Recently, Kane and Wells (KW) proposed a Higgsino-like supersymmetric
(SUSY) particle as a cold-dark-matter candidate \cite{kanewells}. This
proposal was motivated \cite{ambrosanio} by a supersymmetric
interpretation of the CDF $ee\gamma \gamma + \slashchar{E}_T$ event at
Fermilab \cite{CDF} and/or the reported $Z \rightarrow b \bar b$ excess
at LEP.  In general, the lightest SUSY particle (LSP)
is stable (if $R$-parity is conserved), and the other (heavier)
SUSY particles eventually decay to the LSP.  KW
suggest the following explanation of the CDF event:
$\tilde e^+(\rightarrow e^+ \chi_2) \tilde e^-(\rightarrow e^- \chi_2)$,
followed by the photino-like second-lightest neutralino $\chi_2$
decaying radiatively into the lightest neutralino (and LSP)
$\chi$ and a photon.  [Note that for this interpretation to work,
the gaugino mass parameters must satisfy $M_1 \sim M_2$ rather than 
the gauge unification condition $M_1 =(5/3)\tan^2\theta_W M_2$.]
At present there has been only one such event, so that it is
premature to claim discovery
of supersymmetry.  However, the evidence is certainly
intriguing.  It is therefore interesting to investigate the feasibility
of discovering this particular dark-matter candidate with
existing and forthcoming detectors.
The hope is to either rule out this candidate or to detect
it very soon: the payoff, namely the discovery of supersymmetry
and of the dark matter, would be enormous.

The proposed dark-matter candidate is Higgsino-like:
\begin{equation}
     \chi \sim {\rm sin}\beta \tilde H_d^0 + {\rm cos}\beta 
     \tilde H_u^0 + \delta \tilde Z,
\end{equation}
with $\delta<0.1$.  In the models they consider, the LSP
interactions with light fermions $f$ are due primarily to $Z^0$
exchange and are therefore approximated well by a $\chi\chi \bar
f f$ low-energy effective Lagrangian with a coupling
proportional to $\cos
2\beta$, where $\tan\beta$ is the ratio of Higgs vacuum
expectation values.  The dark-matter phenomenology can therefore be 
parameterized simply by the LSP mass, $m_\chi$, and $\cos2\beta$.
In order for the SUSY explanation of the CDF $ee\gamma\gamma+
\slashchar{E}_T$ event to work, the LSP mass should be in the
range $30\, {\rm GeV} \la m_\chi \la 55\, {\rm GeV}$, and the
radiative decay of
the second-to-lightest neutralino requires $\tan\beta<2$
\cite{ambrosanio}.  The model is consistent with constraints to
the $Z^0$ invisible width \cite{width} if $\tan\beta$ is further
restricted to be even closer to unity, or equivalently,
$\cos2\beta$ closer to zero.  The cosmological
abundance of the LSP is inversely proportional to its
annihilation cross section.  In this model, annihilation occurs
predominantly through an $s$-channel $Z^0$ exchange, so the
annihilation cross section is proportional to $\cos^2 2\beta$
and depends also on the mass.  
In the regime of supersymmetric parameter space studied
prior to the work of Kane and Wells, Higgsinos were not found
to be viable dark-matter candidates because they annihilated
too efficiently in the early Universe to retain a relic
density of any significance.  Here, on the other hand, the
smaller values of tan$\beta$ for this candidate lead to a 
smaller annihilation cross section and hence to a cosmologically
interesting relic abundance (note that there is also no coannihilation
with charginos in this case since the Higgsinos are much lighter
than the charginos).  
The requirement that $\Omega_\chi
h^2\la1$ (which derives from a conservative lower limit of 10 Gyr to
the age of the Universe) fixes $\cos^2 2\beta\ga 0.002$.  
Finally, although the CDF
event does not require it, the $Z\rightarrow b\bar b$ anomaly favors a
value of $m_\chi\la 40$ GeV.  The range of models
considered by KW which satisfy these constraints are
shown in their Fig.~1.  The parameter space is restricted
primarily by their favored range for the relic abundance $0.1
\la\Omega_\chi h^2 \la 0.5$ and by the $Z^0$ invisible
width.  This leaves an irregularly shaped region of favored parameter
space which spans roughly the mass range $30 \, {\rm GeV}\, \la
m_\chi \la 40\, {\rm GeV}$ and $1.05 \la \tan\beta \la
1.4$, or equivalently, $0.002\la\cos^2 2\beta \la 0.11$.

KW studied the prospects for direct detection
\cite{witten} of these 
particles in laboratory dark-matter detectors and found them
promising for next-generation detectors \cite{labdetectors}.
In this paper, we point out that for the
Higgsino-like particle they consider, searches for energetic
neutrinos from LSP annihilation in the Sun \cite{SOS} may be an
equally or more
promising avenue toward detection.  The point is that the
Higgsinos they consider have primarily 
axial-vector (rather than scalar) couplings to nuclei, so they
couple to the spin (rather than the mass) of a nucleus.  The
isotopic fraction of terrestrial nuclei with spin is generally
small.  On the other hand, about 75\% of the mass of the Sun is
composed of nuclei with spin (i.e., protons).  Although
detection of weakly-interacting massive particles (WIMPs) with scalar
interactions is probably more
promising with laboratory detectors, detection of WIMPs with
axial-vector interactions is generally more promising with
astrophysical-neutrino detectors \cite{bernard,jkg,taorich}. For
example, Dirac neutrinos,
which have scalar-like interactions, are currently ruled out as
the Galactic dark matter over most of the plausible mass range by
direct-detection experiments \cite{heidelberg}.  On the other
hand, Majorana neutrinos, which have axial-vector interactions,
are ruled out over a large mass range by null searches at
Kamiokande \cite{kamiokande}, but are quite inaccessible to
direct searches \cite{taorich}.  In fact, Rich and Tao found that
the current bounds to the axial-vector WIMP-nucleon interaction
strength from energetic-neutrino searches were roughly three orders of
magnitude stronger than current limits from direct searches for WIMPs
with masses near 30--40 GeV \cite{taorich}.  In this paper we
consider both indirect and direct 
detection of the newly proposed \cite{kanewells} Higgsino
dark-matter candidate.

In Section II, we calculate the energetic-neutrino rates and discuss
the prospects for indirect detection.  In Section III, we review the prospects
for direct detection.  In Section IV, we discuss the results and make
some concluding remarks.

\section{ENERGETIC-NEUTRINO RATES}

If these particles are indeed present with a halo density
$\rho_\chi \sim 0.3\, {\rm GeV}\,{\rm cm}^{-3}$, then some passing through
the Sun would lose enough energy to be captured.  They then
sink to the core of the Sun, and there build up enough density
to start annihilating with each other.  Among the
annihilation products are ordinary neutrinos which
would be observable in various existing detectors here on 
Earth.  The energies of the neutrinos have a broad distribution
centered roughly at a third of the LSP mass.  The detectors with
data already taken include Kamiokande \cite{kamiokande}, IMB \cite{imb},
MACRO \cite{macro}, Frejus \cite{frejus},
Baksan \cite{baksan}, and those being deployed now include
AMANDA \cite{amanda}, NESTOR \cite{nestor}, and super-Kamiokande.
The best technique for inferring the existence
of these neutrinos 
is as follows:  muon neutrinos interact in the rock outside
of the detector, and give rise to upward-going muons which
can be registered in the detector. [Note that both muon neutrinos
as well as muon antineutrinos are produced by the annihilation; these
give rise to upward-going muons and antimuons.  Both have
been included in all our calculations and estimates,
and we use the words `neutrinos' and `muons' to refer
to the sum of particles and antiparticles.]  The muon-energy thresholds
for IMB and MACRO are roughly 2 GeV, 1.7 GeV for Kamiokande,
and roughly 1 GeV for Baksan.
At present, IMB and Kamiokande constrain the flux of energetic
neutrinos from the Sun with energies $\ga 2$ GeV to be
\begin{equation}
     \Gamma_{\rm det} \la 2.1 \times 10^{-2} {\rm m}^{-2} {\rm yr}^{-1}.
\label{constrainedrate}
\end{equation}
In addition to the muon neutrinos, there is a comparable flux of electron
neutrinos produced by the Higgsino annihilation.  For
the relatively low energies considered here, the efficiency for detection of
electron neutrinos may be comparable to that described
above for muon neutrinos. Thus the
sensitivity to energetic neutrinos from Higgsino annihilation
could be improved if one takes both channels into account.
In this paper we focus on limits on the Higgsino particle
that can be obtained from considering the production of
muons only. 
Note that the Earth is composed primarily of spinless
nuclei, so axially-coupled WIMPs will not be captured, and we
expect no energetic-neutrino signal from the Earth for this
dark-matter candidate.

Calculation of the predicted flux of neutrino-induced muons is
straightforward but lengthy.  It must take into account the
complete capture-rate calculation, which includes the
elastic-scattering cross section and the proper kinematic
factors, and the time scale for equilibration between capture
and annihilation.  The neutrinos will be produced by 
decays of $b$ and $c$ quarks and $\tau$ leptons to which the
LSPs annihilate.  An accurate calculation of the neutrino
spectrum must take into account the branching ratios for
annihilation into various final states, hadronization and slowing of
heavy hadrons, the three-body fermion-decay kinematics, and
slowing and absorption of neutrinos in the Sun \cite{ritz,jerrymarc}.
If the LSP has only an axial-vector
coupling to nuclei, the result for the flux of neutrino-induced
muons (for a local halo density of 0.3 GeV~cm$^{-3}$ and
velocity dispersion of 270 km~s$^{-1}$) can be written (Eq.~(9.55)
in Ref.~\cite{jkg})\footnote{Note that there is a factor of
$\xi(m_\chi)$ missing and the $\tanh(t_\odot/\tau_\odot)$ should
be $\tanh^2(t_\odot/\tau_\odot)$ in Eq.~(9.55) in
Ref.~\cite{jkg}.}
\begin{eqnarray}
     \Gamma_{\rm det} & = & (1.65\times 10^{-4}\,{\rm m}^{-2}\,{\rm
     yr}^{-1})\, \sigma_{40} \nonumber \\
	& \times &\tanh^2(t_\odot/\tau_\odot)\,
     (m_\chi/{\rm GeV})\, S(m_\chi/m_p)\, \xi(m_\chi),
\end{eqnarray}
where $\sigma_{40}$ is the cross section for LSP-proton
elastic scattering due to axial-vector interactions in units of
$10^{-40}$ cm$^2$, $S(m_\chi/m_p)$ (where $m_p$ is the proton
mass) is a kinematic suppression factor, $\tau_\odot$ is
the capture-annihilation equilibration timescale and $t_\odot$
is the age of the Sun, and $\xi(m_\chi)$ is a measure of the
second moment of the neutrino energy distribution.  We now
discuss each of these factors.

In the models we are considering, the Higgsino-quark interaction
is due primarily to $Z^0$ exchange.  The
cross section for scattering from a nucleus of mass $m_N$ with an
unpaired proton is approximated by \cite{kanewells}
\begin{eqnarray}
     \sigma_N & = &{2 m_\chi^2 m_N^2 \over \pi (m_\chi+ m_N)^2 } \, G_F^2
     \cos^2 2\beta \nonumber \\
	 &\times & \lambda^2 J(J+1)\,(\Delta
     d+\Delta s -\Delta u)^2;
\label{elastic}
\end{eqnarray}
for unpaired-neutron nuclei switch $\Delta d$ and $\Delta u$ (our
$\Delta d$ and $\Delta u$ are switched relative to those in KW).
For the LSP-proton elastic
scattering cross section, we take $m_N = m_p$ (proton mass)
and the Lande factor $\lambda^2 J(J+1) = 3/4$ in Eq.~(\ref{elastic}).
Here the $\Delta q$'s are the fraction of spin in the proton
carried by each quark.  Using values from a recent compilation
\cite{smc} which give $\Delta d+ \Delta s -\Delta u = 1.24$,
the LSP-proton elastic scattering cross section
evaluates to $\sigma_{40}\simeq 340 \cos^2
2\beta$.  Using Eqs.~(9.21--22) in Ref.~\cite{jkg}, the kinematic factor 
$S(m_\chi/m_p)$ falls in the range 0.5--0.6 for LSP masses
between 30 and 40 GeV.  Furthermore, the equilibration time
scale is given by (Eq.~(9.8) in Ref.~\cite{jkg})
\begin{equation}
     {t_\odot \over \tau_\odot} = 330 \, \left({ C \over
     \sec^{-1} }\right)^{1/2} \left( {\langle\sigma_A v\rangle
     \over {\rm cm}^3\,
     {\rm s}^{-1} }\right)^{1/2} \left( {m_\chi \over 10 {\rm GeV}} \right)
     ^{3/4},
\label{equilibrationtimescale}
\end{equation}
where the capture rate is (Eq.~(9.19) in Ref.~\cite{jkg}),
\begin{equation}
     C=(1.3\times 10^{25}\,{\rm s}^{-1})\,
     \sigma_{40}\, S(m_\chi/m_p)\,(m_\chi/{\rm GeV})^{-1}.
\end{equation}
Here, $\langle\sigma_A
v\rangle$ is the thermally averaged annihilation cross section
times relative velocity in the limit $v\rightarrow 0$
(i.e., the
$s$-wave contribution).  If the neutralino-quark interaction occurs
predominantly via $Z^0$ exchange, the $v\rightarrow0$ annihilation
cross section is 
\begin{equation}
     \langle \sigma_A v \rangle = {G_F^2 \, \cos^2 2\beta \over 8 \pi}
     \, \sum_f\, c_f \, m_f^2,
\label{annihilationxsection}
\end{equation}
where the sum is over the $\tau$ lepton and $b$ and $c$ quarks to
which the Higgsinos annihilate predominantly, $m_f$ is the fermion
mass, and $c_f$ is a color factor (3 for quarks and 1 for leptons).
Eq.~(\ref{annihilationxsection}) evaluates to $\langle \sigma_A v
\rangle \simeq 4.42\times10^{-27}\, {\rm cm}^3\, {\rm s}^{-1}\, \cos^2
2\beta$.  Inserting our expressions for the annihilation cross section
and capture rate into Eq.~(\ref{equilibrationtimescale}), we get
$t_\odot / \tau_\odot \simeq 2600 \, (m_\chi/35\,{\rm
GeV})^{1/4}\cos^2 2\beta$.  For $\cos^2 2\beta\ga0.002$ and the
Higgsino masses of interest, $t_\odot \ga \tau_\odot \ga 5$, so we may
safely set $\tanh^2(t_\odot/\tau_\odot)$ equal to unity.

Finally, there is $\xi(m_\chi)$, a measure of the second moment
of the neutrino energy distribution.
This depends on the branching ratios for LSP-LSP
annihilation into various annihilation channels.  It is
explicitly given by
\begin{eqnarray}
    \xi(m_\chi)=\sum_F B_F [& 3.47 & \Nzsq_{F,\nu}(m_\chi) \nonumber
    \\ & + & 2.08
    \Nzsq_{F,\bar\nu} (m_\chi)],
\label{xieqn}
\end{eqnarray}
where the sum is over all final states to which the LSPs can
annihilate, and $B_F$ is the branching ratio for annihilation to
each channel, which can be calculated with
Eq.~(\ref{annihilationxsection}).  The $\Nzsq_{F,i}$ is the second
moment of the
energy distribution (scaled by $m_\chi^2$) of neutrino type $i$
{}from final state $F$.  Analytic expressions for $\Nzsq$, which
take into account hadronization, stopping of heavy hadrons, the
three-body decay kinematics, and slowing and absorption of
neutrinos as they pass through the Sun, are given in
Ref.~\cite{jerrymarc}, and these are accurate for the
low-mass Higgsinos considered here \cite{joakim}.  As indicated
in Eq.~(\ref{xieqn}), $\xi(m_\chi)$ depends on the annihilation
branching ratios, and the range of possible values is indicated
in Fig.~33 in Ref.~\cite{jkg}.  For LSPs with
masses in the range 30--40 GeV, $\xi$ takes on its largest value
($\sim0.11$) for annihilation into $\tau$ leptons ($B_\tau=1$)
and its smallest value ($\sim0.034$) for annihilation into $b$
quarks ($B_b=1$).  If, as assumed by KW,
annihilation occurs via the $Z^0$ and $\tan\beta$ is near unity,
then $B_F\propto c_f m_f^2$ [c.f., Eq.~(\ref{annihilationxsection})].
Therefore, Higgsinos should
annihilate primarily to $b$ quarks, so we will take
$\xi(m_\chi)=0.034$.  It should be kept in mind, however, that
there will always be some nonzero annihilation branch into
$\tau$ leptons, and in some models, if annihilation via $t$- and
$u$-channel exchange of a stau is larger, it may be significant.
More generally, one can approximate $\xi$ for arbitrary
annihilation branching ratios by noting that annihilation always
occurs almost entirely to $\tau$ leptons and $b$ and $c$ quarks
(since the other quarks are so much lighter).  Furthermore,
the value one would obtain for $\xi$ for the case of
annihilation predominantly into $c$ quarks (i.e., if for
some reason $B_c=1$) is quite close to the value one obtains for the
most likely case of annihilation predominantly to $b$ quarks
(i.e., when $B_b=1$), especially at low energies.
Therefore, for LSPs with masses near 35 GeV, and for arbitrary
branching ratios to $\tau$ leptons and $b$ and $c$ quarks, we can write $\xi
\simeq 0.11 \,B_\tau + 0.034\,(1-B_\tau)$.
Therefore, by taking $B_\tau=0$, we are using a
conservative lower limit for $\xi$, and it could conceivably be
a factor of 3 larger.

Putting together all the factors, the detection rate is given by
\begin{eqnarray}
     \Gamma&&_{\rm det} \simeq  (2.7\times 10^{-2}\,{\rm
     m}^{-2}\,{\rm yr}^{-1})\, (m_\chi/35\, {\rm GeV})\, \cos^2
     2\beta \nonumber \\
	&&\times \left( { \Delta d +\Delta s -\Delta u \over 1.24}
	\right)^2 [3.2\, B_\tau + (1-B_\tau)] 
	 \left[ {S(m_\chi/m_p)
     \over 0.55} \right] \nonumber \\
	&& \times  \tanh^2[2600 \, (m_\chi/35\,{\rm
     GeV})^{1/4}\cos^2 2\beta],
\label{indirectrate}
\end{eqnarray}
where we have included the dependence on the model parameters
$m_\chi$ and $\cos^2 2\beta$, spin content of the proton,
on the annihilation branch to $\tau$ leptons, the kinematic factor,
and the equilibration timescale, although as indicated above, the
dependence on these last two factors will be very weak.
Eq.~(\ref{indirectrate}) is obtained assuming no muon
energy thresholds (i.e. all muons can be detected);
this assumption is a good approximation for detectors with
thresholds near a few GeV (e.g., MACRO, Kamiokande, and
Baksan), since these energies are negligible compared with
the Higgsino mass.  However, for detectors with higher
thresholds (e.g., AMANDA and NESTOR), a good fraction of the
signal may be below threshold.  

The relic density is inversely proportional to $\cos^2 2\beta$.
Since the count rate scales as cos$^2 2 \beta$, for a given mass
the rate drops with increasing $\Omega_\chi h^2$.  Furthermore,
for a given relic abundance, the rate drops with increasing mass
(see Fig.~1 in KW for the dependence of tan$\beta$
on mass and abundance.)
As an example, for $\Omega_\chi h^2 = 0.3$, the predicted count
rates are $\Gamma_{\rm det} \sim 2.6 \times 10^{-3}$ for $m_\chi =
30$ GeV and $\Gamma_{\rm det} \sim 1.1 \times 10^{-3}$ for $m_\chi
= 35$ GeV.

In general, if, as is likely, annihilation occurs predominantly to $b$ quarks,
then the predicted rates fall roughly one to three orders of magnitude
below currently published limits for the $\cos^2 2\beta$ range of
interest, $0.002 \lesssim {\rm cos}^2 2\beta \lesssim 0.11$.  
However, the accumulated exposure of Baksan is greater than
that of Kamiokande, so even better sensitivities (perhaps by a factor
of two) have probably already been achieved.  Also, by performing an
analysis of the data which takes into account the predicted angular
and energy distribution of the neutrino-induced muons, one should be
able to improve the sensitivity with existing data \cite{edsjo}.  More
significantly, the MACRO collaboration expects to reach a sensitivity a
factor of ten or so better than current bounds within
the next five years as it continues to take data.  

Future detectors such as super-Kamiokande, AMANDA, and NESTOR
may have exposures orders of magnitude larger than current
detectors.  Whether they can improve on current sensitivities to
light Higgsinos will depend on the thresholds of these
detectors.  These future detectors may
have much larger thresholds (e.g. 10--30 GeV) so that
the calculations described above would have to be redone.
Although the neutrino energy distribution is
centered roughly at a third the LSP mass, the probability of
detecting a neutrino is proportional to the square of the
neutrino energy.  Therefore, the upward-muon signal is due
to a large extent to the high-energy tail of the neutrino
distribution.  If so, there may still be a significant signal
even for thresholds as high as 10--30 GeV, depending on the
model parameters.  
In addition, for larger thresholds,
the annihilation channel to tau particles via intermediate
staus becomes important because the neutrinos produced
via this channel are stiffer and hence capable of being
above these larger thresholds.

\section{PROSPECTS FOR DIRECT DETECTION}

Let us now briefly review the prospects for direct detection in a
laboratory detector---$^{73}$Ge detector, for example \cite{kanewells}.
The rates 
are controlled by the cross section for Higgsino elastic scattering
{}from a $^{73}$Ge nucleus.  (Note that the  scattering rate from the naturally
abundant isotope,  $^{76}$Ge, is very small since this
isotope has no spin.)  
The cross-section for scattering from a nucleus with an
unpaired neutron, such as $^{73}$Ge, was given in Eq.~(\ref{elastic})
(with $\Delta d$ and $\Delta u$ reversed).
In the single-particle shell model, the Lande factor for
$^{73}$Ge evaluates to $\lambda^2 J(J+1)\simeq0.3$.  However, it
should be kept in mind that the odd-group model predicts a number 80\%
smaller and more detailed calculations suggest it may be 2\%
\cite{ted} to 40\% \cite{dimitrov} smaller than in the single-particle
shell model.  Using the simplest (and most optimistic) value,
Eq.~(\ref{elastic}) evaluates to $\sigma_{73}^{40}\simeq 6.5\times10^4
\cos^2 2\beta [\lambda^2 J(J+1)/0.3]$, in units of $10^{-40}$
cm$^2$, for Higgsinos with masses 30--40 GeV.

In order of magnitude,
the rate for scattering from $^{73}$Ge is $R\sim f_{73}\sigma_{73} \rho
v / (m_\chi
m_N)$ where $\rho$ is the local halo density, $v$ is the halo
velocity dispersion, and $f_{73}$ is the isotopic fraction of
$^{73}$Ge in the sample.  A careful calculation must include the velocity
distribution (and its yearly modulation) of halo dark-matter
particles incident on the detector
and the proper form-factor suppression for spin-dependent scattering
{}from $^{73}$Ge \cite{ted,dimitrov}.  For a Higgsino of mass 40 GeV,
the event rate (including all relevant physical effects) for
scattering in $^{73}$Ge may be obtained from the differential event
rate for detection of an axially-coupled WIMP plotted in Fig.~22 in
Ref.~\cite{jkg}.  The result (averaged over the yearly
modulation) for scattering in natural germanium ($f_{73}=0.078$) is 
\begin{eqnarray}
     R &\simeq & (1.2\times 10^{-5}\,{\rm kg}^{-1}\, {\rm day}^{-1}) \,
	f_{73}\,\sigma_{73}^{40} \nonumber
     \\ & \simeq& 0.0624\, \cos^2 2\beta\, [\lambda^2 J(J+1)/0.3]
\label{directrate}
\end{eqnarray}
This result agrees (well within the
nuclear-physics uncertainties) with the results shown in Fig.~2 in KW,
although our numbers are slightly smaller.
Eq.~(\ref{directrate}) is obtained assuming no thresholds.
However, finite energy thresholds in realistic experiments will
cut out a significant fraction of events and lower the predicted
detection rate accordingly.

Now consider, for illustration, the Cryogenic Dark Matter Search
(CDMS)
experiment \cite{cdms}, which
will first run with 1 kg of natural germanium.  After background
rejection from demanding ionization-calorimetry coincidence, there
will still be a background event rate of roughly $R_b \simeq 1\,
{\rm kg}^{-1}\, {\rm day}^{-1}$.  After a one-year exposure ($E=$ 365
kg-days), the $3\sigma$ sensitivity of the experiment will be roughly
$S\simeq 3\sqrt{R_b/E} \simeq 0.16\, {\rm kg}^{-1}\, {\rm day}^{-1}$.
They also plan to run a similar experiment with roughly 0.5 kg of
enriched $^{73}$Ge.  Assuming the same background-event rate, this
would improve the sensitivity to scattering
{}from $^{73}$Ge to 0.017 kg$^{-1}$ day$^{-1}$ when compared with the
prediction above [c.f., Eq.~(\ref{directrate})] for scattering in
natural germanium.  

For $0.002\la \cos^2 2\beta \la 0.11$, the predicted rate for Higgsino
scattering in natural germanium is 0.0001--0.007 kg$^{-1}$ day$^{-1}$
(again, this is only slightly smaller than the results shown in Fig.~2
of KW).  Therefore, even the most optimistic models (with the
most optimistic nuclear-physics and energy-threshold
assumptions) seem to fall roughly a factor of two below this
forecasted CDMS sensitivity.

\section{DISCUSSION}

The predicted rates for both direct [Eq.~(\ref{directrate})] and
indirect [Eq.~(\ref{indirectrate})] detection of
axially-coupled WIMPs are proportional to the WIMP-nucleon
coupling---in this case, $\cos^2 2\beta$.  For a given WIMP mass, 
we can therefore compare the forecasted
enriched-$^{73}$Ge CDMS sensitivity with the current upward-muon
limit.  Doing so, we find that the enriched-$^{73}$Ge sensitivity
(0.017 kg$^{-1}$ day$^{-1}$) will improve on the {\it current}
limit to the upward-muon flux ($2.1\times10^{-2}$ m$^{-2}$ yr$^{-1}$)
roughly by a factor of 4.  When we compare this with the forecasted
factor-of-ten improvement expected in MACRO, it appears that the
sensitivity of indirect-detection experiments looks favorable.  Before
drawing any conclusions, however, it should be noted that the
sensitivity in detectors with other nuclei with spin may be
significantly better.
We therefore conclude that the two
schemes will be competitive for detection of axially-coupled WIMPs.
Realistically, we must also emphasize that the forecasted
sensitivities of these current experiments will probe only the most
optimistic region of KW's favored parameter space.  It will require
much larger low-background laboratory detectors or
astrophysical-neutrino observatories to probe a good fraction of the
interesting light-Higgsino models.

We should re-emphasize that our estimates for both direct- and
indirect-detection rates may actually be
conservative.  If our halo is flattened (as halos of many
spirals seem to be), then the local halo density could be twice
as large.  If the LSP has a considerable annihilation branch to
$\tau$ leptons, the neutrino rates could be up to a factor of 3 larger.
It should also be kept in mind that---although KW
argued that they should be small---the  LSP-quark interaction
will have some squark-exchange contributions in addition to the
$Z^0$ contributions.  This will probably expand the viable
parameter space.  It also implies that there may realistically
be additional contributions to the axial-vector coupling to
nuclei, and there may be some scalar interaction.  This could enhance
rates for both direct and indirect detection, although much more
dramatically for direct detection.  It would also result in a neutrino
signal from the center of the Earth.

There are also uncertainties which might reduce the rates.  Variations
within current experimental constraints in the spin content of the
nucleon could either increase or decrease both direct- and
indirect-detection rates, perhaps dramatically.  The Lande factor is
likely to be smaller than the value from the single-particle shell
model which we used, and if so, the direct-detection rates will be
lowered accordingly.  There may be sizeable errors in the form factor
for spin-independent scattering which would affect the direct rates.
Finally, we have assumed 100\% direct-detection efficiency.  However,
for the Higgsino mass range of interest, a significant fraction of the
recoils could be below the detection threshold, and the sensitivity
to an LSP passing through the detector would be degraded accordingly.

Within the next five years, detectors at LEP and Fermilab will be able
to confirm or rule out the $ee\gamma\gamma + \slashchar{E}_T$ event that
motivated this Higgsino dark-matter candidate.  Much of the
available parameter space for the SUSY interpretation will
in fact be tested within one year.
If the SUSY interpretation of this event is indeed correct,
then this Higgsino particle exists, but one still does
not know how long it lives; the fact that it escapes the
detector only proves that its lifetime is longer than $\sim 10^{-8}$
seconds.  Proving that this particle is in fact the dark matter
in the halo of our Galaxy would require that it be detected either
directly or indirectly due to its annihilation in the Sun.  Of course
these techniques may be used to rule out the existence of the particle
instead.  In this paper, we have suggested that the timescale for
indirect detection may be comparable to that for experimental
verification or disproof of the motivating event.

To conclude, we have provided estimates of the rates for
indirect detection of Higgsino dark-matter candidates motivated
by collider data.  Our calculations take
into account all relevant physical effects; the accuracy of our
estimates should
be well within the irreducible astrophysics and particle-physics
uncertainties inherent in any such calculation.  Our
results---obtained assuming only a $Z^0$-exchange contribution
to the LSP-quark interaction---suggest that indirect detection 
may provide a realistic alternative avenue toward verification or
falsification of this Higgsino dark-matter candidate.

\acknowledgments

We would like to thank C. Akerib, E. Diehl, R. Gaitskell,
G. Kane, C. Kolda, and G. Tarle for helpful discussions.  MK
thanks the Theory Division at CERN for hospitality,
and MK and KF thank the Fermilab Theoretical Astrophysics Center
(where part of this work was completed) for hospitality.  This work
was supported at Columbia by the D.O.E. under contract DEFG02-92-ER
40699, NASA under NAG5-3091, and the Alfred P. Sloan Foundation,
and at Michigan by NSF-PHY9407194.  Portions of this work (KF)
were completed at the Aspen Center for Physics.

\vfil\eject

\end{document}